# Observation of magnetic adatom-induced Majorana vortex and its hybridization with field-induced Majorana vortex in an iron-based superconductor

Peng Fan,[1,2,†] Fazhi Yang,[1,2,†] Guojian Qian,[1,2,†] Hui Chen,[1,2,3,†] Yu-Yang Zhang,[1,2,4] Geng Li,[1,2,4] Zihao Huang,[1,2] Yuqing Xing,[1,2] Lingyuan Kong,[1,2] Wenyao Liu,[1,2] Kun Jiang,[1,5] Chengmin Shen,[1,2,4] Shixuan Du,[1,2,4] John Schneeloch,[6] Ruidan Zhong,[6] Genda Gu,[6] Ziqiang Wang,[5,*] Hong Ding,[1,3,4,*] and Hong-Jun Gao[1,2,3,4,*]

[1]Beijing National Laboratory for Condensed Matter Physics and Institute of Physics, Chinese Academy of Sciences, Beijing 100190, China

[2]School of Physical Sciences, University of Chinese Academy of Sciences, Beijing 100190, China

[3]Songshan Lake Materials Laboratory, Dongguan, Guangdong 523808, China

[4]CAS Center for Excellence in Topological Quantum Computation, University of Chinese Academy of Sciences, Beijing 100190, China

[5]Department of Physics, Boston College, Chestnut Hill, Massachusetts 02467, USA

[6]Condensed Matter Physics and Materials Science Department, Brookhaven National Laboratory, Upton, New York 11973, USA

[†]These authors contributed equally to this work.
[*]Correspondence to: wangzi@bc.edu, dingh@iphy.ac.cn, hjgao@iphy.ac.cn

**Abstract**

Braiding Majorana zero modes is essential for fault-tolerant topological quantum computing. Iron-based superconductors with nontrivial band topology have recently emerged as a surprisingly promising platform for creating distinct Majorana zero modes in magnetic vortices in a single material and at relatively high temperatures. The magnetic field-induced Abrikosov vortex lattice makes it difficult to braid a set of Majorana zero modes or to study the coupling of a Majorana doublet due to overlapping wave functions. Here we report the observation of the proposed quantum anomalous vortex with integer quantized vortex core states and the Majorana zero mode induced by magnetic Fe adatoms deposited on the surface. We observe its hybridization with a nearby field-induced Majorana vortex in iron-based superconductor FeTe$_{0.55}$Se$_{0.45}$. We also observe vortex-free Yu-Shiba-Rusinov bound states at the Fe adatoms with a weaker coupling to the substrate, and discover a reversible transition between Yu-Shiba-Rusinov states and Majorana zero mode by manipulating the exchange coupling strength. The dual origin of the Majorana zero modes, from magnetic adatoms and external magnetic field, provides a new single-material platform for studying their interactions and braiding in superconductors bearing topological band structures.

**Introduction**

The band structure of iron-based superconductor FeTe$_{0.55}$Se$_{0.45}$ has a nontrivial Z$_2$ topological invariant and supports helical Dirac fermion topological surface states (TSS)[1, 2], which was confirmed recently by spin-resolved and angle-resolved photoemission spectroscopy[3]. Remarkably, below the bulk transition temperature $T_c$, superconducting (SC) TSS were observed with a pairing gap[3]. This makes FeTe$_{0.55}$Se$_{0.45}$ a novel single-material platform for generating Majorana zero modes (MZMs) at the ends of a vortex line[2, 4, 5]. Recently, strong evidence of MZMs inside the magnetic field induced vortices have been observed by scanning tunneling microscope/spectroscopy (STM/S) on the vortex lattice of this type-II superconductor[6-9].

This Majorana platform also presents new challenges for the basic understanding of defect excitations in superconductors with a topological nontrivial band structure, and new possibilities for creating MZMs under different physical conditions. In general, superconductors can host two kinds of defect excitations as in-gap bound states: the Yu-Shiba-Rusinov (YSR) states[10-17] localized at a magnetic impurity and the Caroli-de-Gennes-Matricon (CdGM) states[18-26] inside a magnetic vortex core. To date, these excitations have appeared distinctly in nature; a magnetic impurity induces the YSR states carrying spin, whereas an external magnetic field creates vortices of the whirling supercurrents. In stark contrast to the traditional YSR states, robust zero-bias peaks (ZBPs), sharing the quintessential spectroscopic properties of MZMs, were observed by STM/S at the magnetic interstitial Fe impurities in $FeTe_{0.55}Se_{0.45}$[27], but without applying a magnetic field. A recent theoretical proposal[28] attributes the observed ZBP to a MZM bound to a quantum anomalous vortex (QAV) nucleated spontaneously at the magnetic Fe atom. The role of the magnetic field is played by the exchange coupling of the spin and orbital moment of the Fe impurity located at the $C_4$ symmetric sites, which generates circulating supercurrents by the spin-orbit coupling and modulates the phase of the superconducting order parameter. When the exchange coupling is strong enough, a condition favored by the very small Fermi energy (~ 5meV) in $FeTe_{0.55}Se_{0.45}$[3], a transition from the vortex-free YSR states to the QAV was predicted to take place[28]. A MZM emerges inside the QAV core from the superconducting TSS, since the Berry phase of the Dirac fermions transforms the total angular momentum quantum number of the CdGM vortex core states[18] into integers, naturally supporting a zero-energy bound state[8, 28].

## Results

**Characterizations of samples and Fe adatoms.**

To probe the remarkable nature of defect excitations in the superconducting TSS, we deposit magnetic Fe adatoms on the cleaved (001) surface of a single crystal of FeTe$_{0.55}$Se$_{0.45}$ (Fig. 1a-b) with the substrate temperature below 20 K. Before depositing the Fe adatoms, we scan the surface to ensure that there is no interstitial Fe adatom (See Supplementary Note 1), to avoid the mix of interstitial Fe adatoms and deposited Fe adatoms. In contrast to the growth-induced interstitial Fe impurity in the bulk, the adsorbed Fe adatoms are distributed at various heights above the surface and in different planar locations with respect to the C$_4$ symmetry sites. As a result, the magnetic Fe adatoms with varying exchange couplings reveal much richer phenomena of the defect excitations of the superconducting TSS.

The STM image of a surface region after the atomic deposition (Fig. 1c) shows scattered Fe adatoms as the bright spots with a coverage of ~0.04%. The zero-energy dI/dV map in the same area of Fig. 1c, displays relatively high density of states at the locations of the Fe adatoms (Fig. 1d), consistent with the physical picture that magnetic Fe impurities generate in-gap states. From the statistics of more than one hundred measurements (Supplementary Note 1and Supplementary Fig. 2), we identify two types of in-gap states localized around the Fe adatoms with distinct dI/dV spectra exemplified in Fig. 1e. The type-I adatoms, which represent about 10% of our measurements, exhibit a sharp ZBP reminiscent of a MZM coexisting with other in-gap states in the dI/dV spectrum. In contrast, the conductance spectrum of the type-II adatoms, which represent about 90% of our measurements, shows the YSR states, featuring a pair of in-gap states at particle-hole symmetric peak energy positions, but with asymmetric peak weights. In comparison, the typical dI/dV spectrum on the clean surface without the adatom shows a hard superconducting gap without in-gap states (Fig. 1e).

**QAV states in type-I Fe adatoms.**

We begin with the type-I Fe adatoms. A high-resolution topographic image (Fig. 2a) shows an isolated type-I adatom. A circular pattern appears in the zero-energy dI/dV map in the vicinity of the Fe site (Fig. 2b), with the zero-energy intensity center slightly offset from the Fe site. The breaking of the rotation symmetry is likely due to the canting of the magnetic moment of the Fe adatom away from the surface normal and the presence of spin-orbit coupling. The waterfall-like dI/dV spectra (Fig. 2c) and the intensity plot (Fig. 2d) along the red dashed line-cut in Fig. 2a clearly resolve the ZBP and other peaks at nonzero energies inside the superconducting gap. The ZBP persists to about 2.5 nm from the center, indicating the existence of a localized zero-energy state bound to the Fe adatom with the spatial extent comparable to that of the MZM in a magnetic field-induced vortex core in FeTe$_{0.55}$Se$_{0.45}$[6]. To explore the nature of the discretized in-gap states, we extract several dI/dV spectra from Fig. 2c and show them as a stacking plot in Fig. 2e. Doing so accounts for the spatial distributions of the in-gap states and the sample inhomogeneity[6, 7] due to Te/Se alloying, which has proven to be useful for studying the core states of magnetic field-induced vortices in FeTe$_{0.55}$Se$_{0.45}$[8]. The sequence of discretized bound states, including the zero-energy state, is clearly visible as the pronounced peaks labeled by L$_0$, L$_{\pm 1}$, L$_{\pm 2}$. The energy positions of the conductance peaks (L$_n$) are plotted in Fig. 2f. Intriguingly, the average energies (solid lines) of the discrete quantum states bound to the Fe adatom follow closely a sequence of integer quantization $E_n \approx n\varepsilon$, $n = 0, \pm 1, \pm 2, ...$, with the minigap $\varepsilon \sim 1.0$ meV, the same integer sequence (with a minigap $\sim 0.6$ meV) followed by the quantized vortex core states observed recently[8] in magnetic field-induced vortices that host the MZM[6, 7]. The different values of the minigap are caused by the fluctuation of the superconducting gap $\Delta$ and Fermi energy $E_f$ on different surface, due to the inhomogeneous Te and Se atoms in FeTe$_{0.55}$Se$_{0.45}$. More cases are shown in Supplementary Fig. 3 for the type-I Fe adatoms. Such an integer quantized sequence is the hallmark of the CdGM vortex core states of the superconducting TSS in the quantum limit[8, 28]. Our observations thus provide substantiated and compelling evidence that vortex-like topological defect excitations such as the QAVs nucleate spontaneously at the type-I Fe adatoms and the ZBP corresponds to a vortex MZM.

It is necessary to check the temperature and magnetic field dependence of the ZBP, since a vortex MZM would respond differently than vortex-free defect states. The temperature evolution of the ZBP on the Fe adatom turns out to be very similar to that of the MZM in a field-induced vortex[6]. The ZBP intensity of the center spectrum pointed by the black arrow in Fig. 2c decreases with increasing temperatures, and becomes almost invisible at 4.2 K (Fig. 2g). The magnetic field dependence is also very similar to that of a vortex MZM as the ZBP does not split or broaden for fields up to 8 T (Fig. 2h), provided that no field-induced vortices enter the region near the adatom when the magnetic field is applied. The ZBP is however sensitive to the location of the adsorbed Fe adatom. We found that all type-I adatoms producing integer quantized bound states anchored by the ZBP are adsorbed at the high-symmetry sites in the center of four Te/Se atoms. To test the robustness of this finding, we manipulate a type-I adatom by the STM tip to a different location away from the $C_4$ symmetric site. The ZBP disappears and a pair of in-gap state at nonzero energy emerges (Supplementary Note 3 and Supplementary Fig. 4). After annealing the sample to 15 K and performing the measurement again at 0.4 K, the adatom diffuses back to its original high-symmetry site and the ZBP reappears. Thus, the high-symmetry site is a prerequisite for the induced ZBP, which agrees with the proposed theory that the orbital magnetic moment of the Fe adatoms at $C_4$ symmetric locations plays an important role for the nucleation of the QAV[28].

**YSR states in type II Fe adatoms.**

Next, we turn to study the type-II Fe adatoms (Supplementary Note 4). In contrast to the type-I adatoms, the conductance exhibits predominantly a pair of in-gap peaks at nonzero energies without the ZBP. Applying an external magnetic field, we observe that the peaks shift approximately linearly to higher energies (Supplementary Fig. 5d) away from the Fermi level, consistent with a pair of spin-polarized YSR in-gap states. We find that the type-II Fe adatoms are adsorbed at myriad locations on or off the high-symmetry axis and induce YSR states at different energies, indicative of broadly varying exchange couplings to the superconducting quasiparticles. In special cases, we also observe YSR states located very close to zero energy, which nevertheless split under the magnetic field (Supplementary Note 4 and Supplementary Fig. 6) and are therefore distinct from the robust ZBP observed on the type-I Fe adatoms.

**Reversible transition between YSR states and MZMs in some type II Fe adatoms**

These observations motivate us to manipulate the exchange coupling between the magnetic adatoms and the substrate by tuning the tip to sample distance[29,30]. In the approaching-tip process, the STM tip needs to be positioned on top of the Fe adatom, which makes it impossible to acquire spatial maps while keeping the position of the Fe adatom frozen. The electrostatic force of an approaching tip can prod and move the Fe adatom in directions parallel and perpendicular to the surface (Fig. 3a), which can affect the atomic orbital moment of the Fe adatom and the spin-orbit exchange coupling to the superconductor. In STM/S, the tunnel-barrier conductance $G_N \equiv I_t/V_s$ , where $I_t$ is the tunneling current and $V_s$ is the bias voltage, governs the tunnel coupling and changes with the tip-sample distance. Performing tunneling conductance measurements as a function of $G_N$, we find that the energies of the YSR states are modulated ubiquitously when the tip approaches the type-II Fe adatoms (Supplementary Note 5). The observed crossing and reversal of the in-gap states (Supplementary Fig. 7), a trademark of the YSR states, confirms that reducing the tip to sample distance monotonically increases the exchange coupling between the Fe atom and the superconductor.

Unexpectedly, as the STM tip approaches a significant number of the type-II Fe adatoms (~ 27%), the pair of YSR states modulates with increasing $G_N$, but then coalesces into a single ZBP in the water-fall plot of dI/dV spectra (Fig. 3c) and the intensity plot (Fig. 3d), which remains robust under further increase of the barrier conductance. Note that the emergence of the ZBP out of the YSR states is different from the transition point between the screened spin-singlet and doublet ground states[29,31], where the two YSR states are approximately degenerate at zero energy as marked by the red arrow in Fig. 3d. To probe the change in the nature of the in-gap states with different barrier conductance, we repeated the entire process under an applied magnetic field. The dI/dV spectra and the intensity plot obtained under 6 T (Fig. 3e-f) show that the vortex-free YSR states no longer cross zero energy, due to the Zeeman splitting that removes the accidental degeneracy. However, the emergence of the unsplit ZBP at higher $G_N$ is unabridged even at such a high field, indicating that ZBP corresponds to a single MZM robust against an applied magnetic field. This identification is further corroborated by performing the measurements on type-I Fe adatoms that show the ZBP associated with the MZM upholds its integrity and does not shift or split with increasing $G_N$ in a field as high as 6 T (Supplementary Note 6 and Supplementary Fig. 8). The compelling evidence attributes the novel coalescence of in-gaps states toward the ZBP to the change in the nature of the magnetic impurity-induced defect state from the vortex-free YSR state to a vortex state with a vortex MZM (Fig. 3b), which is fully consistent with the theoretical prediction that increasing the exchange coupling of an Fe impurity induces a transition from the YSR states to the QAV states hosting a MZM in FeTe$_{0.55}$Se$_{0.45}$ superconductors[28]. We note that the entire tip-manipulation process is carried out locally without affecting the stability of the superconducting topological surface states. The SC gap in the STM spectrum (Fig. 3) does not close and the transition to the QAV state is reversible. The vortex MZM naturally arises in the QAV core due to the superconducting topological surface states on the surface of FeTeSe[28]. Manipulating the Fe adatom by the approaching tip to the $C_4$ symmetric site and closer to the superconducting surface allows the magnetic Fe to sustain its orbital magnetic moment in addition to the spin moment and increases the spin-orbit exchange coupling, which generates the circulating supercurrents around the defect, leading to the flux trapping and the nucleation of a vortex state. The transition between the YSR states and the MZM is even reversible, as the dI/dV spectra as a function of the barrier conductance retrace that shown in Fig. 3e-f upon withdrawing the tip (Fig. 3g-h) in a controlled manner. The transition is also replicable when the type-II Fe adatom under the tip in Fig. 3 is moved to a different location about 1 nm away (Supplementary Fig. 9). These observations reveal the unprecedented nature of defect excitations in the superconducting TSS where local magnetic moment and screening currents are inextricably connected through the magnetoelectric effect.

**Hybridization between MZMs in QAV and field-induced vortex.**

The phase coherence of the MZMs is stored nonlocally and protected by the topological degeneracy against environmental decoherence caused by local perturbations, which is the central to the idea of topological quantum computing[32,33]. The coupling of two MZMs sufficiently close by annihilates of the nonabelian anyonic zero modes and creates a pair of fermionic states at nonzero energies. This coupling process usually requires two overlapping magnetic field induced vortices, which is difficult to control on the Abrikosov lattice[7, 8, 34]. Our system allows a new possibility, i.e., the hybridization between MZMs hosted in a QAV and a field-induced vortex (Fig. 4). The zero-energy dI/dV map (Fig. 4a) shows a MZM in the QAV nucleated at a type-I Fe adatom in a magnetic field of -0.2 T, also visible in the intensity plot (Fig. 4b) as sharp ZBPs along the line cut across the adatom. A field-induced vortex is observed to enter the field of view subsequently. The latter sits very close to the Fe site, thus enlarges significantly the region with spectral weight at zero-energy (Fig. 4c). Remarkably, acquiring the intensity plot along the same line cut shows that the ZBP splits into two peaks separated by an energy spacing ~ 0.25 meV (Fig. 4d). During the second round of the measurements, the vortex creeps away. The zero-energy map recovers (Fig. 4e) and the ZBP reemerges in the intensity plot (Fig. 4f). Throughout the hybridization process, the temperature and the magnetic field are kept stable and the change in the position of the field-induced vortex is due to the spontaneous vortex creeping. It is necessary to point out that the creeping of the vortices is observed quite often when we detected MZMs in the filed-induced vortices in our previous work. Three representative dI/dV spectra corresponding to the three conditions are extracted and displayed in Fig. 4g for a better comparison. Repeating the measurements on the same Fe adatom in a higher magnetic field of -3 T reveals again the splitting of the ZBP caused by the presence of a nearby field-induced vortex (Fig. 4h), with a larger energy spacing ~ 0.35 meV (Fig. 4i), possible due to the shorter distance and stronger overlap of the two MZMs as indicated by the smaller ring feature in the zero-energy map in Fig. 4h compared to Fig. 4c. Moreover, a field-induced vortex can also enter the field of view without causing detectable splitting of the ZBP bound to the Fe adatom or the vortex MZM when they are relatively far apart (Supplementary Note 8 and Supplementary Fig. 10). These observations further support the identification of the ZBP induced by type-I Fe atoms as the MZM and concurrently provide the first experimental evidence for the hybridization between two vortex MZMs as illustrated in Fig. 4j.

**Discussion**

Our STM/S measurements on magnetic Fe adatoms deposited on the surface of FeTe$_{0.55}$Se$_{0.45}$ superconductors revealed the spontaneous formation of anomalous vortex matter with integer quantized core states and MZMs in zero external field, and the reversible transition between a YSR impurity and a

QAV with increasing exchange interaction strength. These fundamental properties, previously unobserved in superconductors, are consistent with the predications of the theoretical proposal of the QAV[28]. They have broad implications for superconductors with a nontrivial topological band structure and spin-momentum locked superconducting TSS, which have also been found in several other iron-based superconductors such as LiFeAs[35,36], $(Li_{0.84}Fe_{0.16})OHFeSe$[37], and $CaKFe_4As_4$[38]. We also noticed a recent preprint posted on the arxiv[39], where the result shows absence of spin-polarization on a ZBP. It can be included in our results of type II Fe adatoms, where the YSR states can locate at a quantum phase transition point. This result can be calarified by measurements under a high magnetic field. Together with the observed hybridization of the MZMs in the QAV nucleated at the Fe adatom and the nearby field-induced Abrikosov vortex, our findings demonstrate that magnetic adatoms coupled to the superconducting TSS provide a realistic materials platform for creating and studying the interactions between the nonabelian topological vortex MZMs.

**Methods**

**Materials and measurement.** High-quality single crystals of $FeTe_{0.55}Se_{0.45}$ with Tc of 14.5 K are grown using the self-flux method[6]. The samples are cleaved in situ and immediately transferred into a STM head. The Fe adatoms are deposited from a high-purity (99.95%) single crystal Fe rod acquired from *ESPI METALS* to the surface of $FeTe_{0.55}Se_{0.45}$ at a temperature below 20 K. Before Fe deposition, we scan the surface of $FeTe_{0.55}Se_{0.45}$ to ensure that there is no visible interstitial Fe impurity (IFI). The STM experiments are performed in an ultrahigh vacuum ($1 \times 10^{-11}$ mbar) LT-STM systems (USM-1300s-$^3$He), which can apply a perpendicular magnetic field up to 11 T. STM images are acquired in the constant-current mode with a tungsten tip. The energy resolution calibrated on a clean Pb (111) surface is about

0.27 meV. The voltage offset calibration is followed by a standard method of overlapping points of I-V curves. Differential conductance (dI/dV) spectra are acquired by a standard lock-in amplifier at a frequency of 973.1 Hz, under modulation voltage $V_{mod}$ = 0.1 mV. Low temperature of 0.4 K is achieved by a single-shot $^3$He cryostat.

**Data Availability**

All relevant data are available from the corresponding authors upon reasonable request.

**Acknowledgements**

We thank Xi Dai, Shengshan Qin, Shiyu Peng, Sankar Das Sarma for useful discussions. The work at IOP is supported by grants from the National Natural Science Foundation of China (11888101, 61888102 and 11674371), the National Key Research and Development Projects of China (2016YFA0202300, 2018YFA0305800 and 2019YFA0308500), and the Chinese Academy of Sciences (XDB28000000, XDB07000000, 112111KYSB20160061). The work at BNL and BC is supported by grants from US DOE, Basic Energy Sciences (DE-SC0012704, DE-FG02-99ER45747).

**Author contributions**

H.-J. G. and H.D. designed STM experiments. P.F., G.Q. and H.C. performed STM experiments with assistance of F.Y., Z.H., Y.X. and C.S.. J.S., R.Z. and G.G. provided samples. K.J. and Z.Q.W. provided theoretical explanations. P.F., F.Y., G.Q. and H.C. processed experimental data with input from Y.Z., G.L., L.K., W.L. and S.D.. All the authors participated in analyzing experimental data, plotting figures, and writing the manuscript. Z.Q.W., H.D. and H.-J. G. supervised the project.

**Competing interests:**

The authors declare that they have no competing interests.

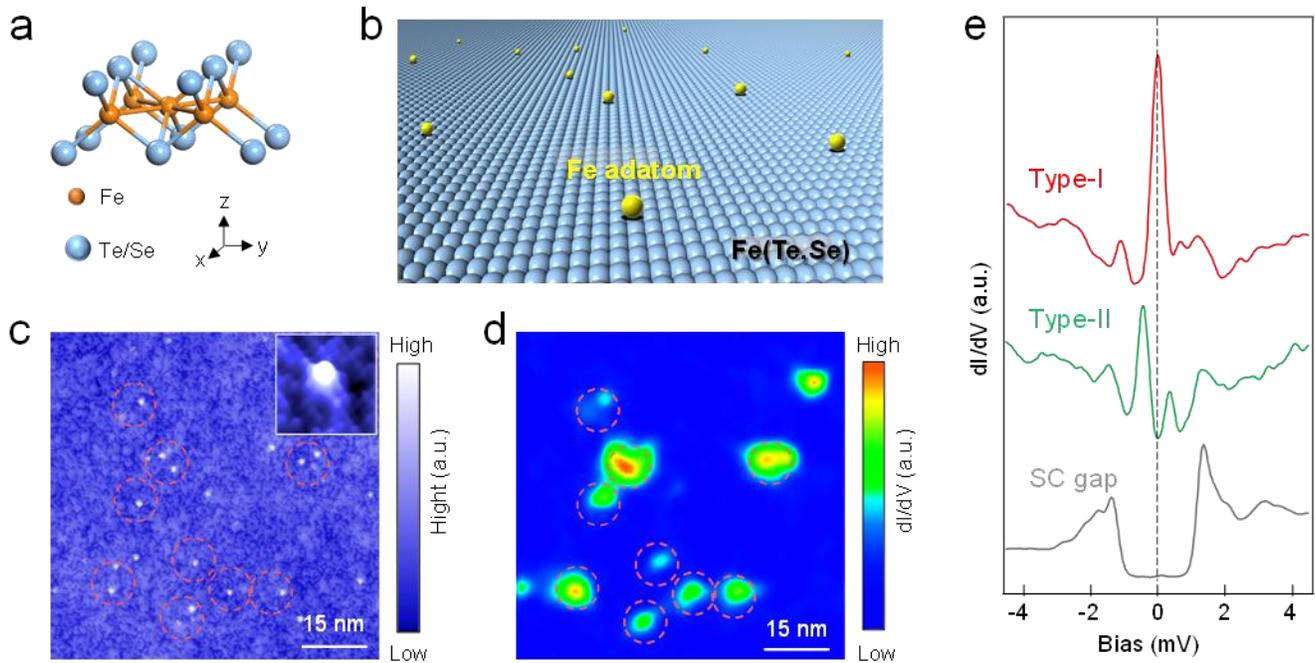

**Fig. 1 Characterization of deposited Fe adatoms on FeTe$_{0.55}$Se$_{0.45}$ surface. a** Crystal structure of Fe(Te,Se). **b** Schematics of Fe adatoms deposited on the surface of FeTe$_{0.55}$Se$_{0.45}$. The adsorption sites are random, including both on and off of the C$_4$ symmetric sites in the center of four Te/Se atoms and at various heights above the surface (Supplementary Note 1). **c** A STM image ($V_s$ = -10 mV, $I_t$ = 100 pA) after atomic Fe atom deposition. The bright dots correspond to Fe adatoms with a coverage of 0.04%. Inset: A 4 nm × 4 nm atomic resolution topographic image showing a single Fe adatom located at a high-symmetry site. **d** A dI/dV map ($V_s$ = -10 mV, $I_t$ = 100 pA) of **c** at zero energy. The high conductance areas and the Fe adatom positions are in good spatial correspondence, as highlighted by the red dashed circles. **e** Typical dI/dV spectra ($V_s$ = -10 mV, $I_t$ = 200 pA) showing the zero bias peak on type-I Fe adatoms (red curve), YSR states (green curve) on type-II Fe adatoms, and the clean superconducting (SC) gap away from the adatoms (gray curve).

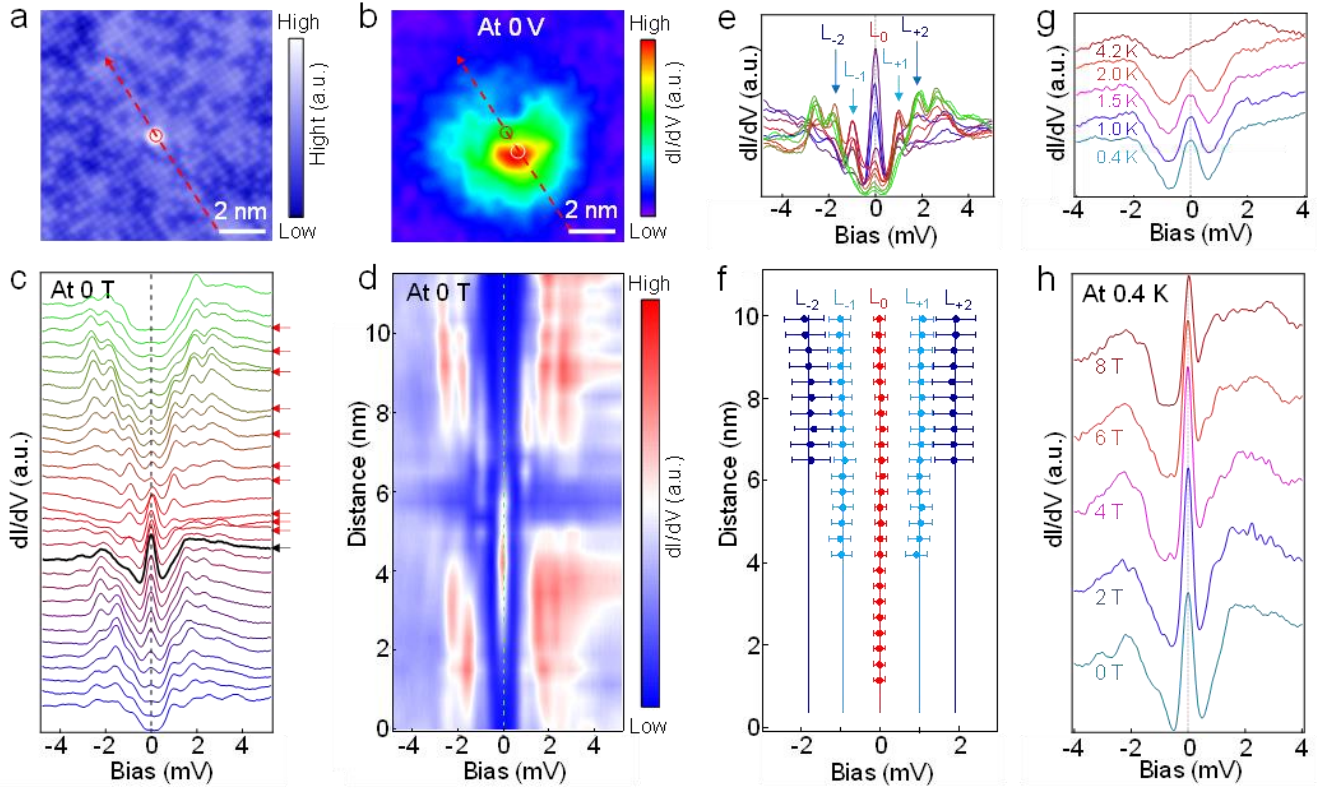

**Fig. 2 ZBP and integer quantized in-gap states on type-I Fe adatoms. a** An atomic-resolution topographic image ($V_s$ = -10 mV, $I_t$ = 100 pA) of a single type-I Fe adatom (red circle). **b** A zero-energy dI/dV map ($V_s$ = -10 mV, $I_t$ = 200 pA) of the same area in **a**. **c** and **d** dI/dV spectra and the corresponding intensity plot ($V_s$ = -10 mV, $I_t$ = 200 pA) along the line-cut indicated by the red dashed and arrowed line in **a** and **b**. The decay length of the ZBP is about 2.5 nm. **e** A stacking plot of several dI/dV spectra ($V_s$ = -10 mV, $I_t$ = 200 pA) marked by the red arrows in **c**. The energies of the in-gap states are labeled as $L_0$, $L_{\pm 1}$, $L_{\pm 2}$. **f** The energies of the in-gap states at different spatial positions along the line-cut. Error bars are the FWHM of a Gaussian fit. The energy values of the solid lines are calculated as the average energy of the in-gap states, showing the same integer quantization as the topological vortex core states anchored by the MZM. **g** Temperature dependence of the dI/dV spectra ($V_s$ = -10 mV, $I_t$ = 200 pA) measured at zero-energy intensity center (white circle in **b**) with the corresponding spectrum (the black curve) pointed by the black arrow in **c**. The ZBP intensity decreases with increasing temperature and becomes indiscernible at 4.2 K. **h** Magnetic field dependence of the dI/dV spectra ($V_s$ = -10 mV, $I_t$ = 200 pA) also acquired at the zero-energy intensity center (white circle in **b**) at 0.4 K. The ZBP remains robustly at zero energy without splitting.

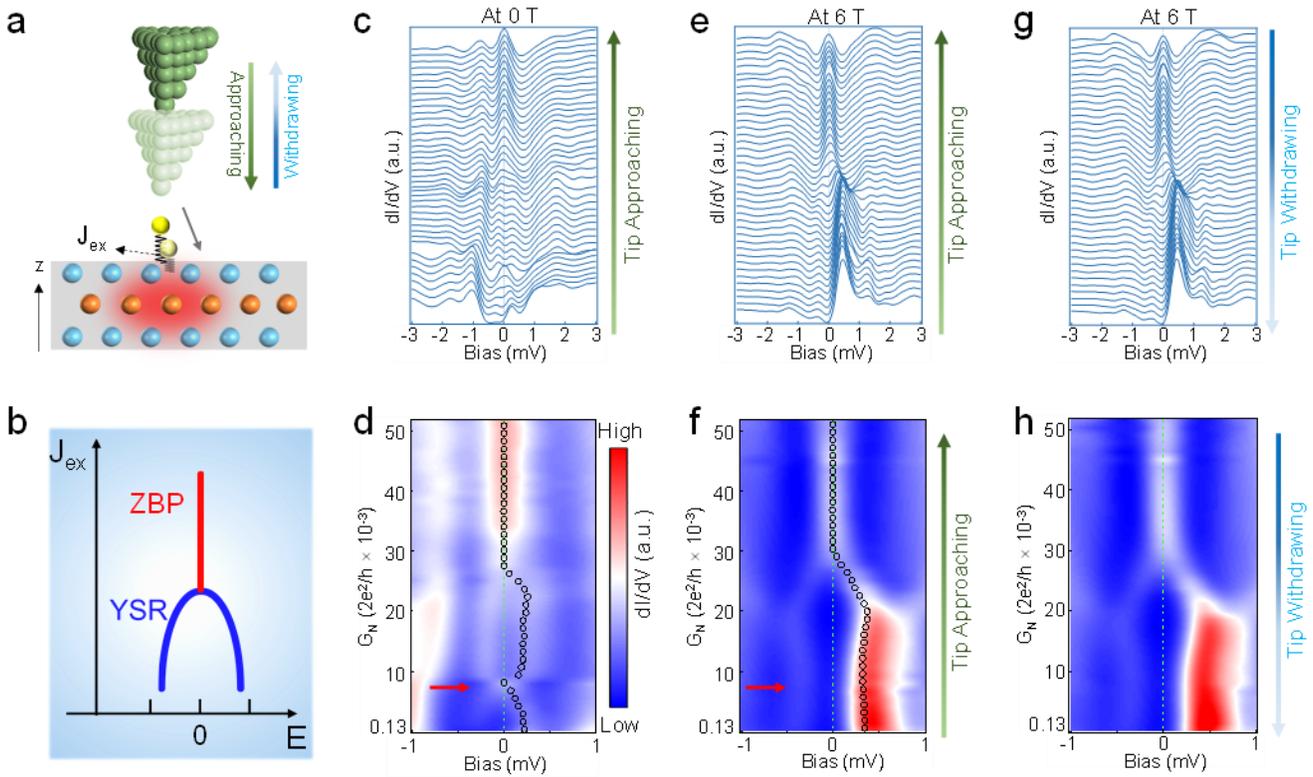

**Fig. 3 Reversible transitions between YSR states and a robust ZBP/MZM induced by modulating the exchange coupling of type-II Fe adatoms using the STM tip. a** Schematics illustrating an approaching STM tip on top of a type-II Fe adatom can move the Fe adatom in directions parallel and perpendicular to the Te/Se surface and modulate the spin-orbit exchange coupling ($J_{ex}$) to the superconductor. **b** Schematic phase diagram from the vortex-free YSR states to an anomalous vortex MZM represented by the ZBP with the increasing exchange coupling $J_{ex}$. **c** and **d** Tunnel barrier conductance dependence of the dI/dV spectra in **c** and its intensity plots in **d**, showing the evolution of vortex-free YSR states into a robust ZBP at 0 T under an approaching tip. Red arrow in **d** indicates the position of an accidental near degeneracy of the YSR states. Black circles in **d** trace the peak positions. **e** and **f** are the same as **c** and **d**, but measured in a magnetic field of 6 T. Black circles in **f** trace the peak positions. The accidental degeneracy of the YSR states in 0 T in d is removed by the 6 T magnetic field, while the evolution to the ZBP remains robust. **g** and **h** are the same as **e** and **f**, but showing the transition from ZBP to YSR states by withdrawing the STM tip at 6 T, indicating the transition is reversible.

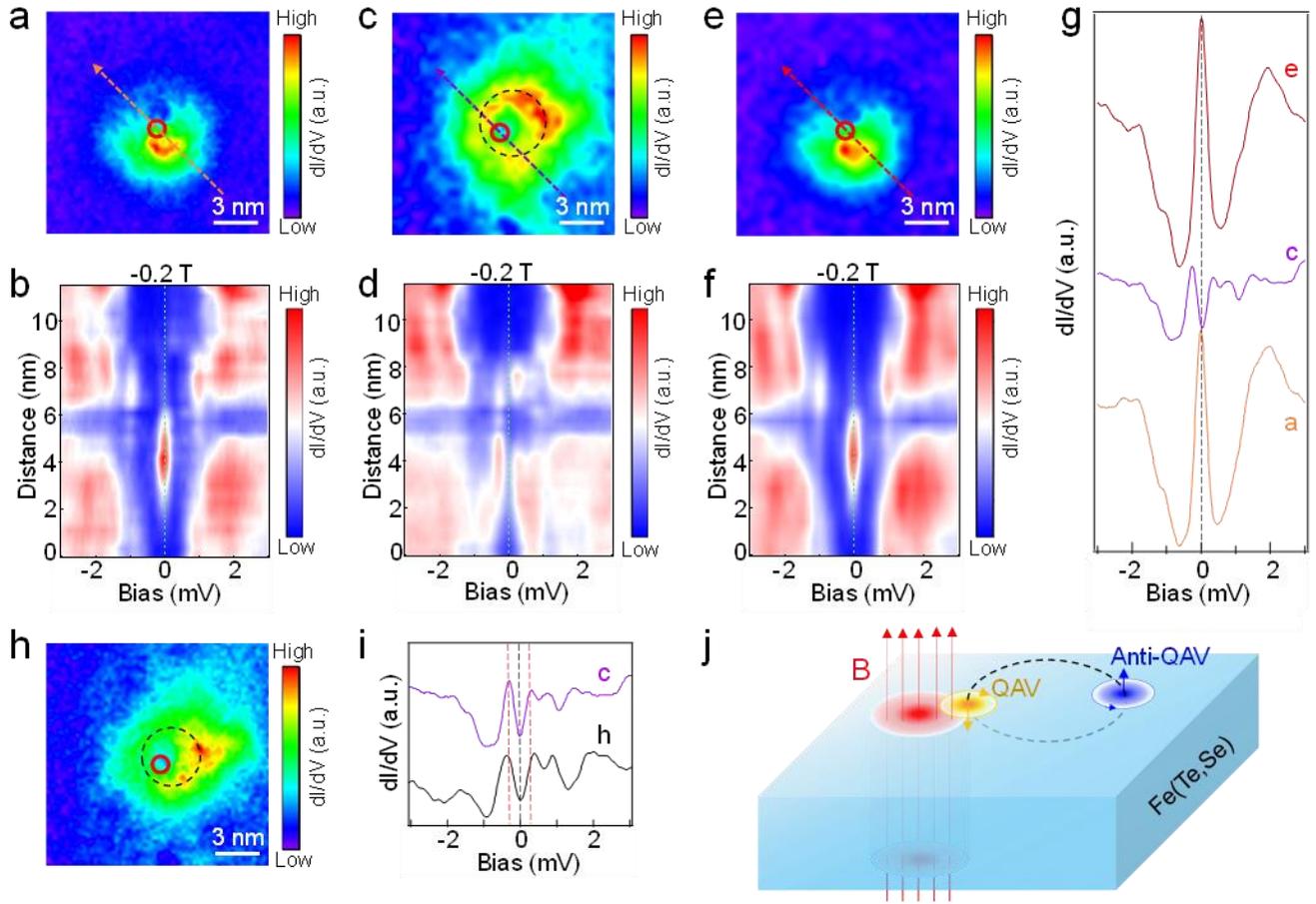

**Fig. 4 Hybridization between two MZMs in the QAV and field-induced vortex. a** Zero-energy dI/dV map ($V_s$ = -10 mV, $I_t$ = 100 pA) around a type-I Fe adatom in a magnetic field of -0.2 T, showing the spatial distribution of the MZM in an isolated QAV nucleated at the adatom. **b** Intensity plot of dI/dV spectra ($V_s$ = -10 mV, $I_t$ = 200 pA) along the line-cut indicated by the arrowed red dashed line in **a**, showing a robust ZBP. **c** and **d** are the same as **a** and **b**, but with an additional field-induced vortex pinned near the same Fe adatom visible in the zero-energy map in **c**. The ZBP disappears in the intensity plot of the dI/dV spectra along the same line-cut in **d** and two in-gap states with an energy level spacing of 0.25 meV emerge. **e** and **f** are the same as **c** and **d**, but after the field-induced vortex creeps away. The MZM returns and the zero-energy dI/dV map and the intensity plot of the dI/dV spectra along the line cut recover those in **a** and **b**. **g** Three dI/dV spectra obtained on the Fe adatom in **a**, **c**, and **e** ($V_s$ = -10 mV, $I_t$ = 200 pA), respectively. **h** Zero-energy dI/dV map around the same Fe adatom at -3T ($V_s$ = -10 mV, $I_t$ = 100 pA), showing an additional field-induced vortex pinned close to the adatom. **i** dI/dV spectra obtained on the Fe adatom in **h** and **c** ($V_s$ = -10 mV, $I_t$ = 200 pA). Both spectra show disappearance of ZBP and emergence of two in-gap states separated by an energy spacing that is larger (0.35 meV) in the case of **h** compared to **c** (0.25 meV). **j** Schematics of the interaction between two MZMs in the QAV and the field-induced vortex.

Supplementary Information for

# Observation of magnetic adatom-induced Majorana vortex and its hybridization with field-induced Majorana vortex in an iron-based superconductor

Peng Fan,[1,2,†] Fazhi Yang,[1,2,†] Guojian Qian,[1,2,†] Hui Chen,[1,2,3,†] Yu-Yang Zhang,[1,2,4] Geng Li,[1,2,4] Zihao Huang,[1,2] Yuqing Xing,[1,2] Lingyuan Kong,[1,2] Wenyao Liu,[1,2] Kun Jiang,[1,5] Chengmin Shen,[1,2,4] Shixuan Du,[1,2,4] John Schneeloch,[6] Ruidan Zhong,[6] Genda Gu,[6] Ziqiang Wang,[5,*] Hong Ding,[1,3,4,*] and Hong-Jun Gao[1,2,3,4,*]

[†]These authors contributed equally to this work.

[*]Correspondence to: wangzi@bc.edu, dingh@iphy.ac.cn, hjgao@iphy.ac.cn

SUPPLEMENTARY NOTE 1: Characterizations of samples and Fe adatoms

Previous measurements of the vortex core states in $FeTe_{0.55}Se_{0.45}$ under an external magnetic field have indicated that there are both topological trivial and topological nontrivial regions on the surface[1]. Before Fe deposition, we scan the surface of $FeTe_{0.55}Se_{0.45}$ to ensure that there is no interstitial iron atom (Supplementary Fig. 1a). We can also detect MZMs in vortices induced by an external magnetic field as exemplified in Supplementary Fig. 1, which demonstrates that the sample surface has topological nontrivial regions. Then, we perform atomic Fe deposition and focus on Fe adatoms. An example of a single Fe adatom is shown in the atomic-resolution STM topography (Supplementary Fig. 2), where the $C_4$ symmetric site is highlighted by the vertices of the bright dashed grid. It can be seen clearly that this Fe adatom is located at the high-symmetry site. Away from the Fe adatom, the "higher" sites correspond to the Te atoms and the "lower" ones to the Se atoms. The line-profile of the Fe adatom yields a height of 1.61 Å and the peak width at half maximum of 1.5 nm (Supplementary Fig. 2b). In general, the adsorption sites of the Fe adatoms can be on or off the $C_4$ symmetry axis piercing vertically through the center of four Te/Se atoms and located at different heights to the surface. From the statistics of hundreds of measurements, we find that the height of deposited Fe adatoms varies from 1.1 Å to 2 Å as shown in the histogram (Supplementary Fig. 2c) with an average value of 1.56 Å that is higher than the typical height (~1 Å) of the interstitial Fe impurities[2] in $FeTe_{0.55}Se_{0.45}$.

SUPPLEMENTARY NOTE 2: Integer quantized in-gap states at type-I Fe adatoms

We show the analysis of three additional type-I Fe adatoms in Supplementary Fig. 3. Similar to the case presented in Fig. 2 of the main text, the zero-energy dI/dV maps (Supplementary Figs. 3a, d and g) and the dI/dV spectra along the line-cuts across the adatoms (Supplementary Figs. 3b, e and h) exhibit sharp ZBPs together with several pairs of in-gap states, which do not disperse spatially. Performing the analysis in Fig. 2 of the main text, we find that the energies of the in-gap states at each adatom follow a series of near-integer quantization, which can be seen from the energies of the peak positions and ratios displayed in Supplementary Figs. 3c, f and i. Similar integer quantization of the in-gap states has been observed in the magnetic field-induced vortices in $FeTe_{0.55}Se_{0.45}$[1], providing compelling evidence that these in-gap states are the topological CdGM vortex core states of the QAV[3] nucleated at type-I Fe adatoms on the superconducting TSS, and the ZBP corresponds to the MZM.

SUPPLEMENTARY NOTE 3: Manipulation of a single type-I Fe adatom

To study the role of the high-symmetry adsorption site of the type-I Fe adatoms, we manipulate the location of the Fe adatom characterized in Fig. 2 of the main text using the STM tip. We first move the Fe adatom off the $C_4$ symmetric center surrounded by four Te/Se atoms into a different location (Supplementary Fig. 4a). Then we measure the zero-energy dI/dV map and the dI/dV spectra cross the adatom site along a line-cut as shown in Supplementary Fig. 4b and 4c. The pattern in the zero-energy dI/dV map and the dI/dV spectra along the line-cut change significantly compared to Fig. 2b-c in the main text: the ZBP disappears and a pair of in-gap states emerge at non-zero energies. Subsequently, we heat the sample up to 15 K and observe that the Fe atom diffuses back to the original high-symmetry site Supplementary Fig. 4d. We then cool the sample back down to 0.4 K and perform again the dI/dV measurements. As can be seen in Supplementary Fig. 4e-f, the ZBP reappears and the zero-energy dI/dV map as well as the spectra along the line-cut recover the spatial distribution of the zero-mode and the conductance spectra of those before the manipulation (Fig. 2b-c in the main text). The results indicate that the high-symmetry site is necessary for the induced ZBP by the type-I Fe adatoms. Finally, we remove the Fe adatom using the STM tip (Supplementary Fig. 4g). The measured spectra show a hard superconducting gap in the same area, demonstrating that the in-gap states originate from the Fe adatom (Supplementary Fig. 4h-i).

**SUPPLEMENTARY NOTE 4: YSR states and Zeeman splitting on type-II Fe adatoms**

The high-resolution topographic image shows an isolated type-II adatom (Supplementary Fig. 5a). The dI/dV spectra along the red dashed line-cut in Supplementary Fig. 5a are displayed in the waterfall plot (Supplementary Fig. 5b) and the intensity plot (Supplementary Fig. 5c). In contrast to the type-I adatoms, the conductance exhibits predominantly a pair of in-gap peaks at nonzero energies without the ZBP. Applying an external magnetic field, we observe that the peaks shift to higher energies (Supplementary Fig. 5d) away from the Fermi level, consistent with a pair of spin-polarized YSR in-gap states. The energy positions of the YSR states under different magnetic fields can be fitted by a linear Zeeman splitting with the g-factor of about 0.88 (Supplementary Fig. 5e).

Because the energies of the in-gap states are controlled by the exchange interaction strength, it is known that accidental near degeneracy of the YSR states close to zero-energy (the near-zero YSR states) can arise. The atomic resolution STM topography (Supplementary Fig. 6a) and the zero-energy dI/dV map

(Supplementary Fig. 6b) show an example of such a type-II Fe adatom. The evolution of the dI/dV spectra along the red line-cut across the adatom (Supplementary Fig. 6c) and its intensity plot (Supplementary Fig. 6d) clearly reveal the near-zero energy bound states whose spectroscopic features are similar to the ZBP detected on the type-I Fe adatoms in the absence of an external magnetic field. However, when external magnetic field is applied, the accidental degeneracy is removed as the near-zero YSR states show apparent Zeeman splitting into the usual YSR states (Supplementary Fig. 6e). The energy splitting can be well fit by a linear function in the magnetic field with a g-factor about 0.5. These observations clearly demonstrate that one can distinguish the near-zero YSR states from the robust ZBP by applying external magnetic fields. The near-zero YSR states with Zeeman splitting account for about 6.4% of the Fe adatoms in our measurements.

## SUPPLEMENTARY NOTE 5: Modulating YSR states with approaching STM tip at type-II Fe adatoms

The method of varying the tip-sample distance has been used to change the coupling between the adatoms and the surface. In recent works[4, 5], the exchange coupling between a magnetic impurity and a BCS superconductor (Pb) has been shown to be tunable and the quantum transition of the YSR states by approaching the STM tip at the impurity sites has been realized. We first check the effectiveness of this method in our system by approaching the tip at the type-II Fe adatom site shown in Supplementary Fig. 7a. We measure the tip-sample distance by the offset $z$ in Supplementary Fig. 7b, which shows that reducing $z$ leads to a monotonic increase in the tunnel barrier conductance $G_N$. Before reducing the tip-sample distance, i.e. at the distance offset $z = 0$ in Supplementary Fig. 7b, the particle-hole symmetric (in energy) YSR states have a larger spectral weight on the negative energy side, visible in the dI/dV spectra at the bottom in Supplementary Fig. 7c as well as in the intensity plot (Supplementary Fig. 7d) at the smallest tunnel barrier conductance $G_N$. As the STM tip is made to approach the adatom, the YSR states evolves with increasing $G_N$ toward and cross zero energy to the superconducting gap edge with the larger spectral weight shifted to be on the positive energy side (Supplementary Fig. 7c-d). This is consistent with the increasing of the exchange coupling between the Fe adatom and the superconducting surface that causes a quantum transition of the YSR states with the increasing tunnel-barrier conductance $G_N$. Note that the *z*-offset decreases smoothly with the barrier conductance $G_N$ (Supplementary Fig. 7b), indicating no change of the tip and the Fe atom during the approaching process.

## SUPPLEMENTARY NOTE 6: Robustness of the ZBP under approaching STM tip at type-I Fe adatoms

We carry out tests on the robustness of the ZBP induced by type-I Fe adatoms under an approaching STM tip. The normalized STM intensity plots in zero-field and under an external magnetic field of 6T are shown in Supplementary Figs. 8c and 8d as a function of the tunnel barrier conductance $G_N$. The ZBP at the type-I Fe adatom is remarkably robust and does not shift or split with increasing $G_N$. The robustness of the ZBP at type-I Fe adatoms against increasing exchange interaction is illustrated in the schematic diagram in Supplementary Fig. 8b, which should be contrasted to the transition from the YSR states to the ZBP observed at the type-II Fe adatoms illustrated in Fig. 3b in the main text.

## SUPPLEMENTARY NOTE 7: Replicable transition between YSR states and a vortex MZM

Here we show that when the type-II adatom under the STM tip in Fig.3 of the main text is moved to another location about 1 nm away (Supplementary Fig. 9a), the transition from the YSR states into a ZBP as shown in Figs. 3b-d at the original location is replicated (Supplementary Fig. 9b). About 27.3% of the type-II Fe adatoms studied by the tip-approaching measurements exhibit the transition from the YSR states to the zero energy bound state tied to the ZBP. This phenomenon should not be induced by the saturation of the exchange coupling strength during the tip approaching, in which case the final energy value should not remain at zero energy frequently. There can be several reasons for the rest of the type-II Fe adatoms not to produce a topological vortex with an MZM. The approaching tip in these cases may be unsuccessful at nudging the adatoms to the high-symmetry adsorption sites, and/or the spin-orbit exchange interaction has not been increased sufficiently for the nucleation of the QAV. These type-II Fe adatoms may be located in the nontopological surface regions where the topological surface states are absent locally, which is in-line with the observation that only about 20% of the field-induced vortices show the MZM on the same sample surface[1, 6]. Moreover, based on the QAV theory[3], the nucleation of an anomalous vortex at one Fe adatom on the surface is accompanied by that of an anomalous antivortex at another Fe adatom, such that the magnetic flux lines are continuous. While experimental test for this would require multiple STM/S tips in the future, we note that the type-II Fe adatoms displaying the YSR states to vortex MZM transition tend to be located in regions with other Fe adatoms, which can be more favorable for the nucleation of a quantum anomalous vortex-antivortex pair.

**SUPPLEMENTARY NOTE 8: Coexistence of MZMs in the QAV and the field-induced vortex**

We also observe a field-induced vortex hosting an MZM in an external magnetic field of 1T, which coexists with the MZM in the QAV nucleated at the type-I Fe adatom. The zero-energy dI/dV map (Supplementary Fig. 10a) shows that the field-induced vortex core center is about 7 nm away from the Fe adatom, which is larger than the ones studied in Fig. 4c and 4h of the main text. The intensity plot of the dI/dV spectra (Supplementary Fig. 10a) along the line-cut across both the field-induced vortex and the QAV clearly reveals two ZBPs localized separately in the cores of both vortices. When the external magnetic field is switched off, the field-induced vortex disappears while the QAV at the adatom remains in the zero-energy map (Supplementary Fig. 10c). The dI/dV spectra along the same line-cut show that the hard superconducting gap recovers where the field-induced vortex was located, but the robust ZBP in the QAV nucleated at the Fe adatom. The results indicate that this surface area is in the topological region and there is an MZM in the field-induced vortex in Fig. 4c and Fig. 4h of the main text. These findings support the observation of hybridization between the two MZMs inside the Fe adatom induced topological QAV and the field-induced topological Abrikosov vortex presented in Fig. 4 of the main text.

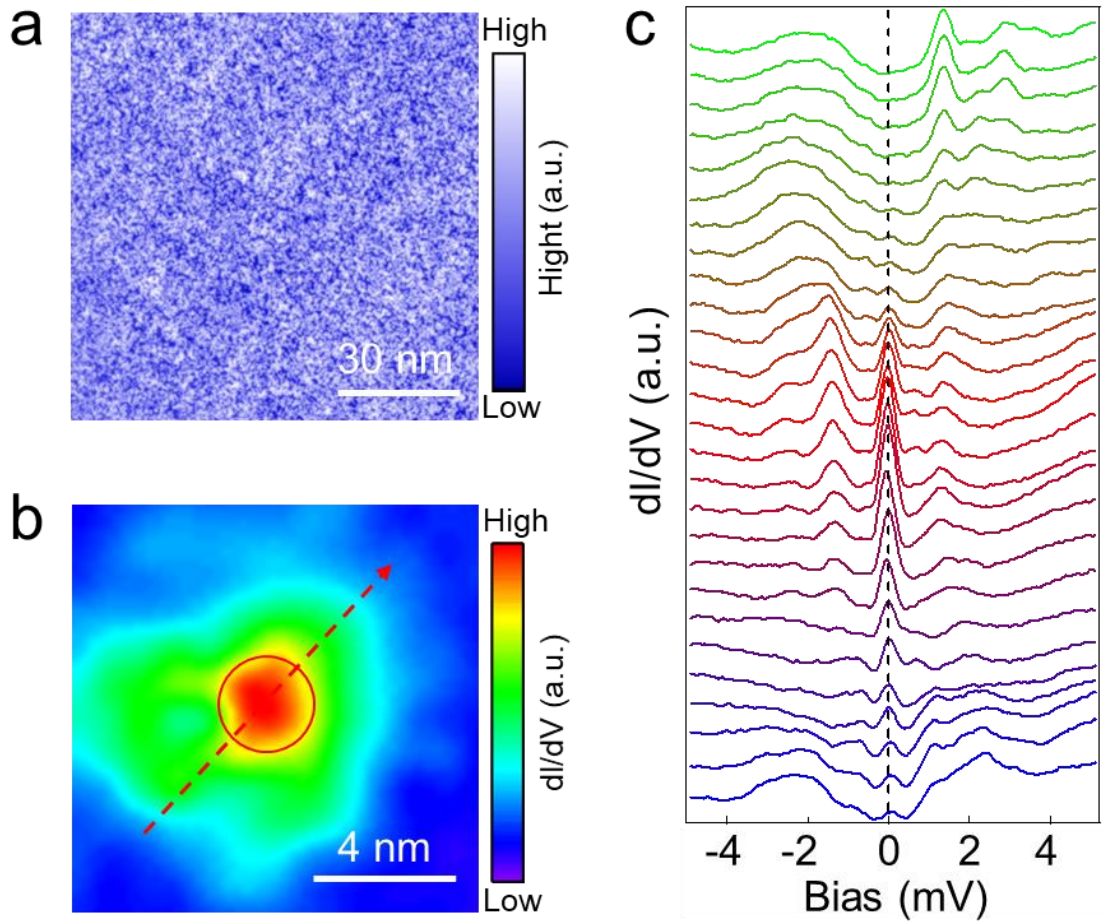

**Supplementary Figure 1. MZM in a field-induced vortex detected in FeTe$_{0.55}$Se$_{0.45}$. a**, A STM image ($V_s$ = -10 mV, $I_t$ = 100 pA) of FeTe$_{0.55}$Se$_{0.45}$ before depositing Fe adatoms. There is no interstitial iron atom. **b**, A zero-energy dI/dV map ($V_s$ = -10 mV, $I_t$ = 100 pA) of a field-induced vortex with a MZM in the same area of **a**. The field-induced vortex center is marked by the red circle in **a** and **b**. **c**, dI/dV spectra ($V_s$ = -10 mV, $I_t$ = 200 pA) along the line-cut indicated in **b**, showing a sharp ZBP that does not split or shift with the changes of spatial position, consistent with previous observation of vortex MZMs in the same material.

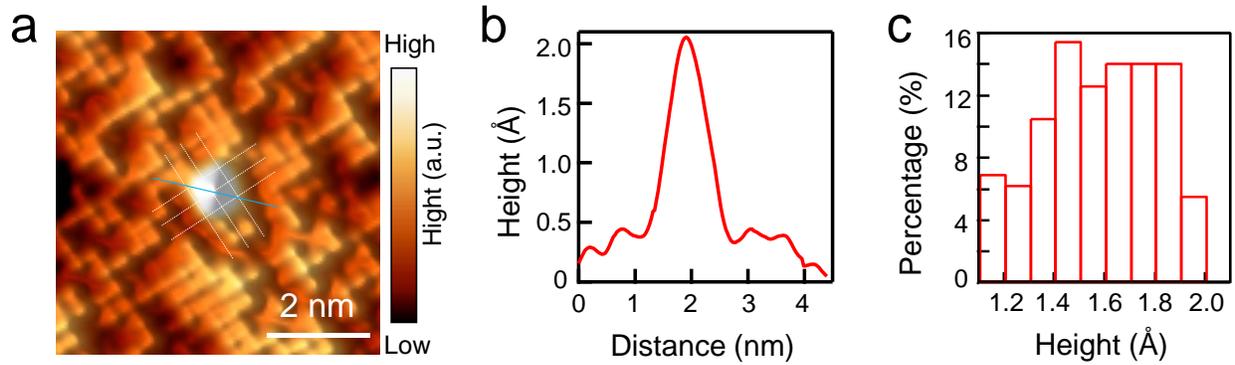

**Supplementary Figure 2. Characterization of deposited single Fe adatom on FeTe$_{0.55}$Se$_{0.45}$ surface.** **a**, A high-resolution STM image ($V_s$ = -10 mV, $I_t$ = 100 pA) showing a single Fe adatom located at the C$_4$ symmetric site. The C$_4$ symmetric site is highlighted by the vertices of the white dashed grid. **b**, A line-profile ($V_s$ = -10 mV, $I_t$ = 100 pA) along the light blue line in **a**, showing that the single Fe adatom is 1.61 Å in height and 1.5 nm in width (defined as the peak width at half maximum). **c**, A histogram showing the statistics of the height distribution of hundreds of Fe adatoms. The height varies from 1.1 Å to 2.0 Å.

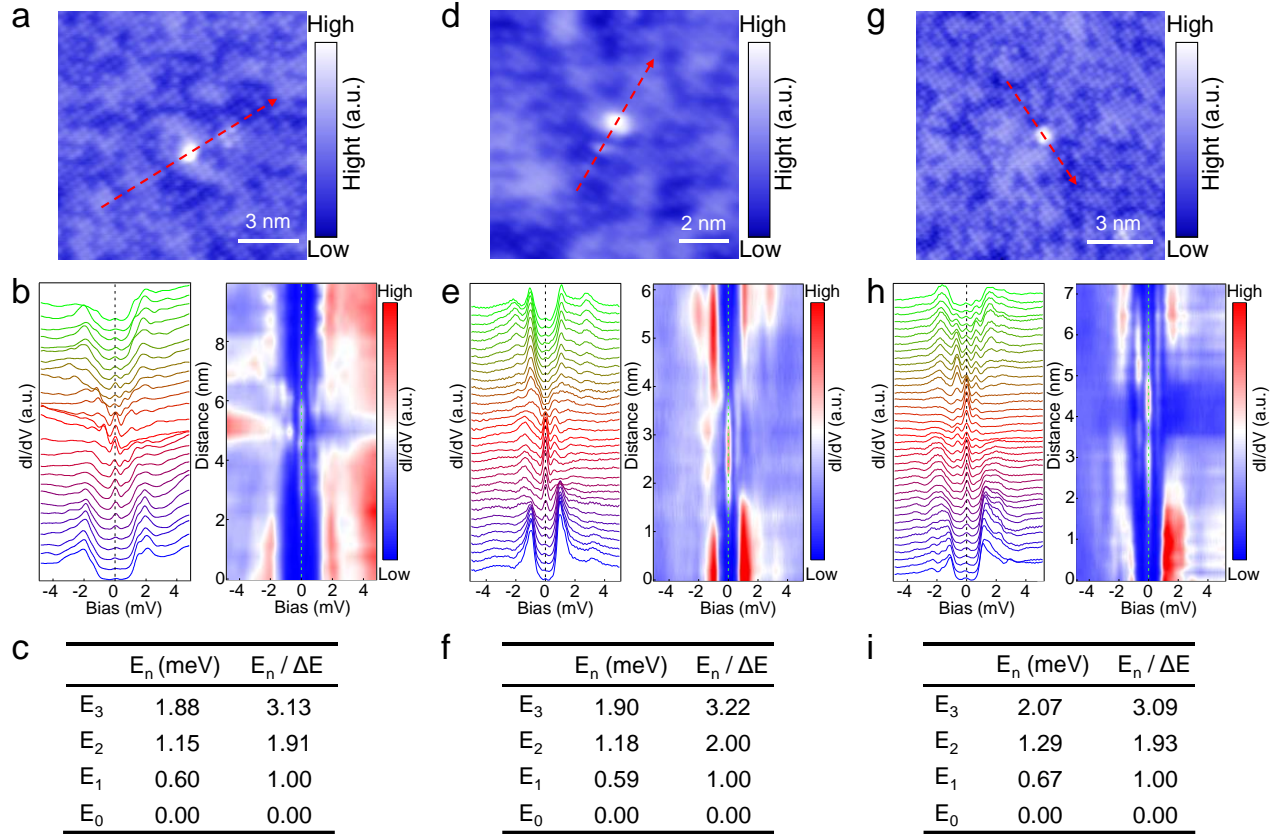

**Supplementary Figure 3. Integer quantized in-gap states on type-I Fe adatoms. a**, **d** and **g**, Atomic-resolution topographic images ($V_s$ = -10 mV, $I_t$ = 100 pA) of type-I Fe adatoms. **b**, **e** and **h**, Waterfall and intensity plots of dI/dV spectra ($V_s$ = -10 mV, $I_t$ = 200 pA) measured along the line-cut indicated correspondingly in **a**, **d** and **g** across the three adatoms, showing a sharp ZBP coexisting with several pairs of discrete in-gap states. **c**, **f** and **i**, Lists of the energy positions of the in-gap states corresponding to **b**, **e** and **h**, respectively, showing near integer quantized energy level spacing which is the hallmark of the CdGM vortex core states of the topological vortex.

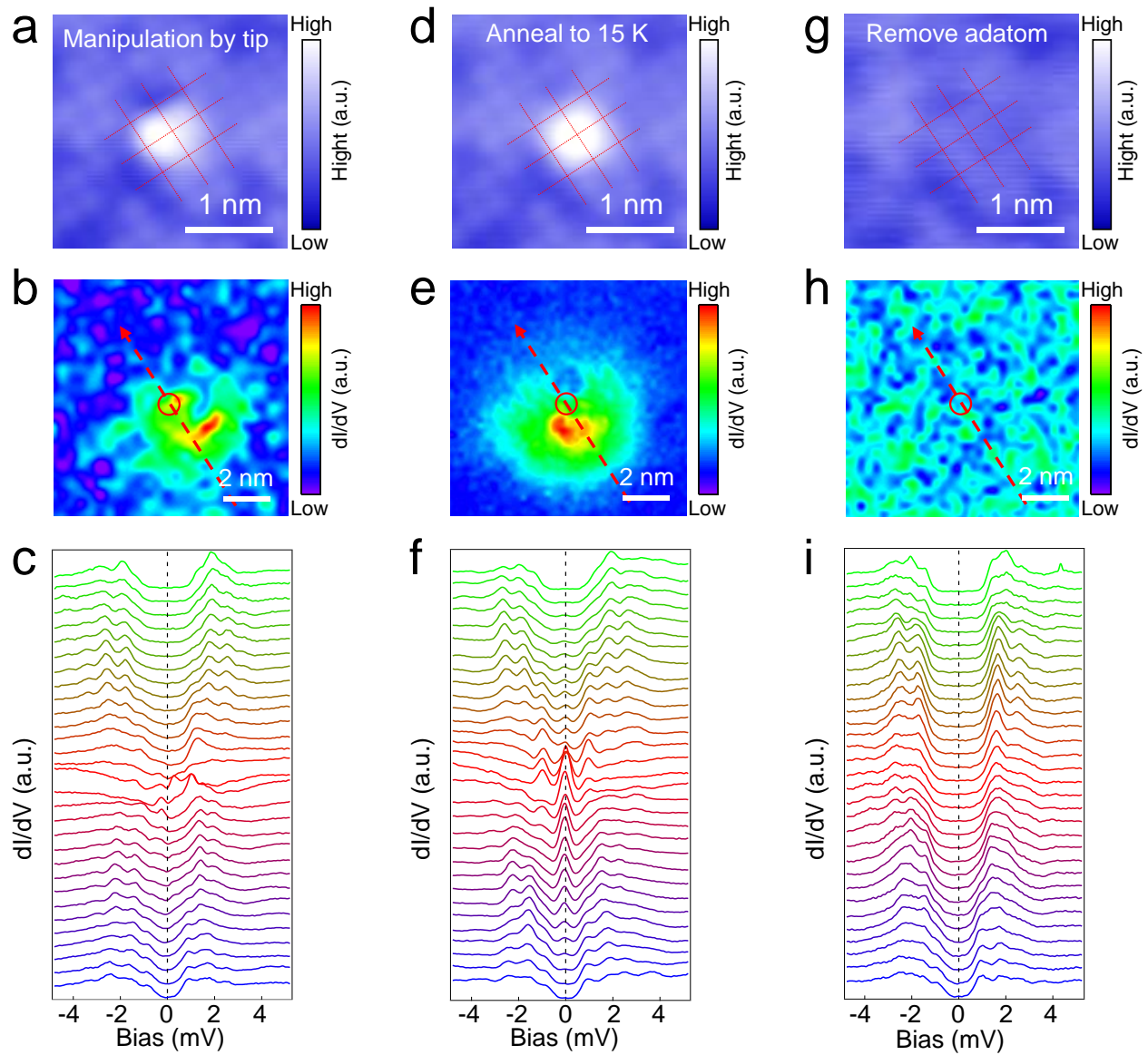

**Supplementary Figure 4. Manipulation of a type-I Fe adatom. a**, A STM image ($V_s$ = -10 mV, $I_t$ = 100 pA) showing that the Fe adatom in Fig. 2 of the main text has been moved off the C$_4$ symmetric center by the STM tip. **b**, A zero-energy dI/dV map ($V_s$ = -10 mV, $I_t$ = 100 pA) around the Fe adatoms in **a**. The pattern is significantly different from that before the manipulation shown in Fig. 2b of the main text. **c**, dI/dV spectra ($V_s$ = -10 mV, $I_t$ = 200 pA) along the line-cut indicated in **b** showing that the ZBP disappears and other in-gap states emerge. **d**, A STM image ($V_s$ = -10 mV, $I_t$ = 100 pA) showing that the Fe adatom moves back to the center of four Te adatoms by *in situ* annealing the sample and cooling back to 0.4 K. **e**, A zero-energy dI/dV map ($V_s$ = -10 mV, $I_t$ = 100 pA) around the Fe adatoms in **d**. The pattern recovers

the spatial distribution of the zero-mode before the manipulation (Fig. 2b of the main text). **f**, dI/dV line-cut spectra ($V_s = -10$ mV, $I_t = 200$ pA) along the red dashed arrow in **e** showing that the ZBP reappears. **g**, A topographic image ($V_s = -10$ mV, $I_t = 100$ pA) showing the removal of the Fe adatom from the field of view by the STM tip. **h**, The zero-energy dI/dV map ($V_s = -10$ mV, $I_t = 100$ pA) around the Fe adatoms in **g** shows that there are no low energy bound states after removing the Fe adatom. **i**, dI/dV line-cut spectra ($V_s = -10$ mV, $I_t = 200$ pA) along the red dashed arrow in **h**, showing the evolution of a hard superconducting gap.

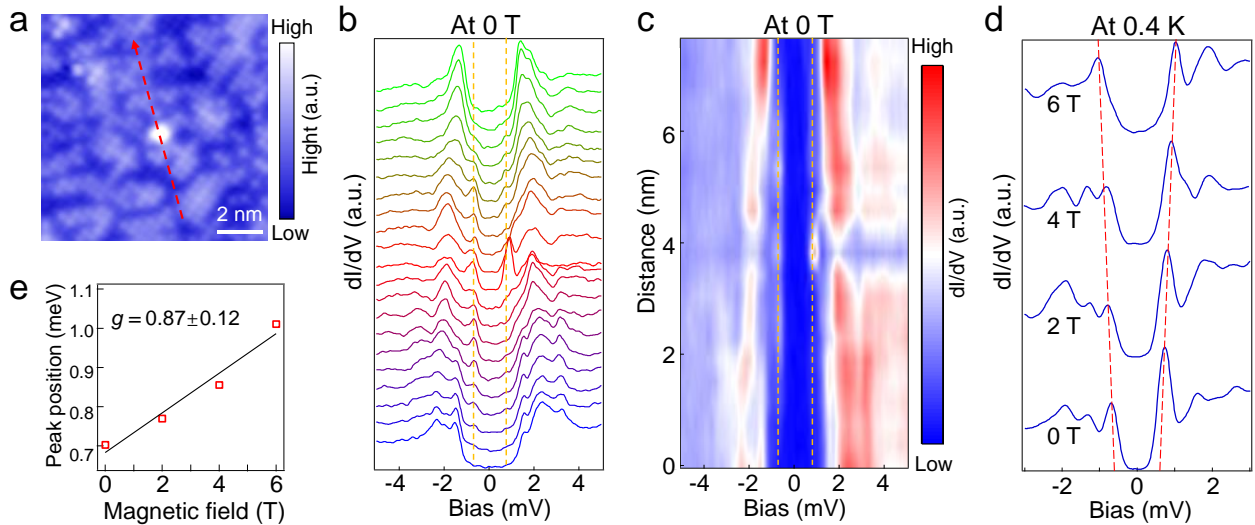

**Supplementary Figure 5. YSR states on type-II Fe adatoms and Zeeman splitting. a**, An atomic-resolution topographic image ($V_s$ = -10 mV, $I_t$ = 100 pA) showing a single type-II Fe atom. **b** and **c**, dI/dV spectra ($V_s$ = -10 mV, $I_t$ = 200 pA) and the corresponding intensity plot ($V_s$ = -10 mV, $I_t$ = 200 pA) along the line-cut indicated by the red dashed and arrowed line in **a**. The dashed lines highlight the emergence of the pair of YSR states near the Fe adatom. **d**, Magnetic field dependence of the dI/dV spectra ($V_s$ = -10 mV, $I_t$ = 200 pA) acquired at the adatom. Red dashed lines trace out of the spin-polarized Zeeman splitting of the YSR states. **e**, A linear fit of the YSR state energy as a function of magnetic field, yielding an estimated *g*-factor ~ 0.87.

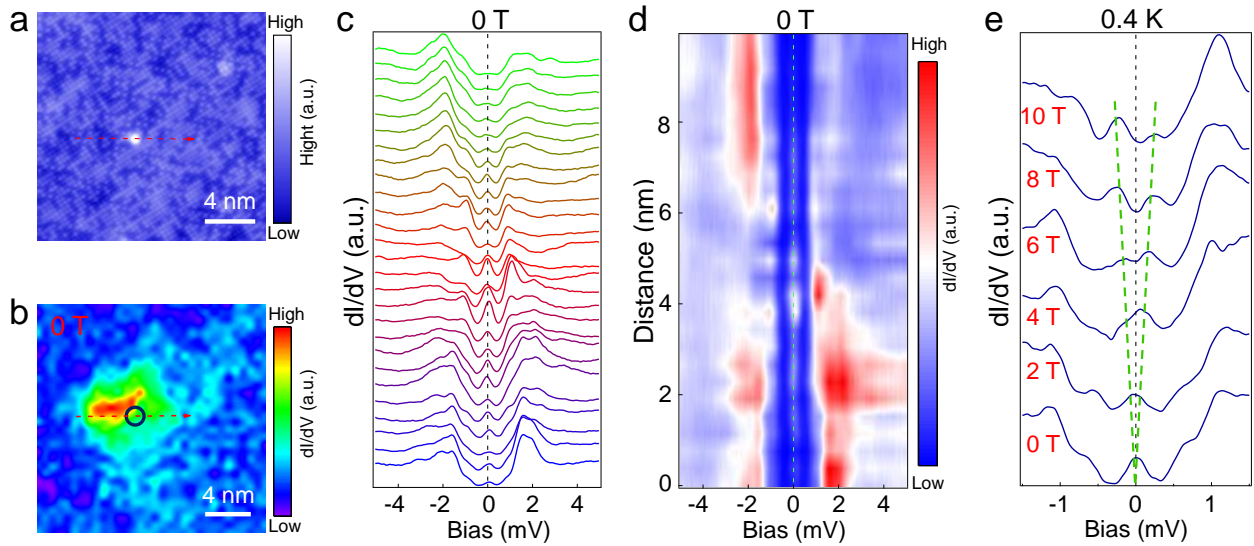

**Supplementary Figure 6. Near-zero energy YSR states and Zeeman splitting. a**, An atomic-resolution STM image ($V_s = -10$ mV, $I_t = 100$ pA) of a type-II Fe adatom. **b**, A zero-energy dI/dV map ($V_s = -10$ mV, $I_t = 100$ pA) of the same area in **a**. **c** and **d**, dI/dV spectra ($V_s = -10$ mV, $I_t = 200$ pA) and the corresponding intensity plot along the line-cut indicated by the red dashed arrow in **a**, showing nearly degenerate YSR states close to zero energy highlighted by the dashed line. **e**, Magnetic field dependence of the dI/dV spectra ($V_s = -10$ mV, $I_t = 200$ pA) acquired at the adatom site (black circle in **b**). The green dashed lines trace out of the Zeeman splitting of the YSR states.

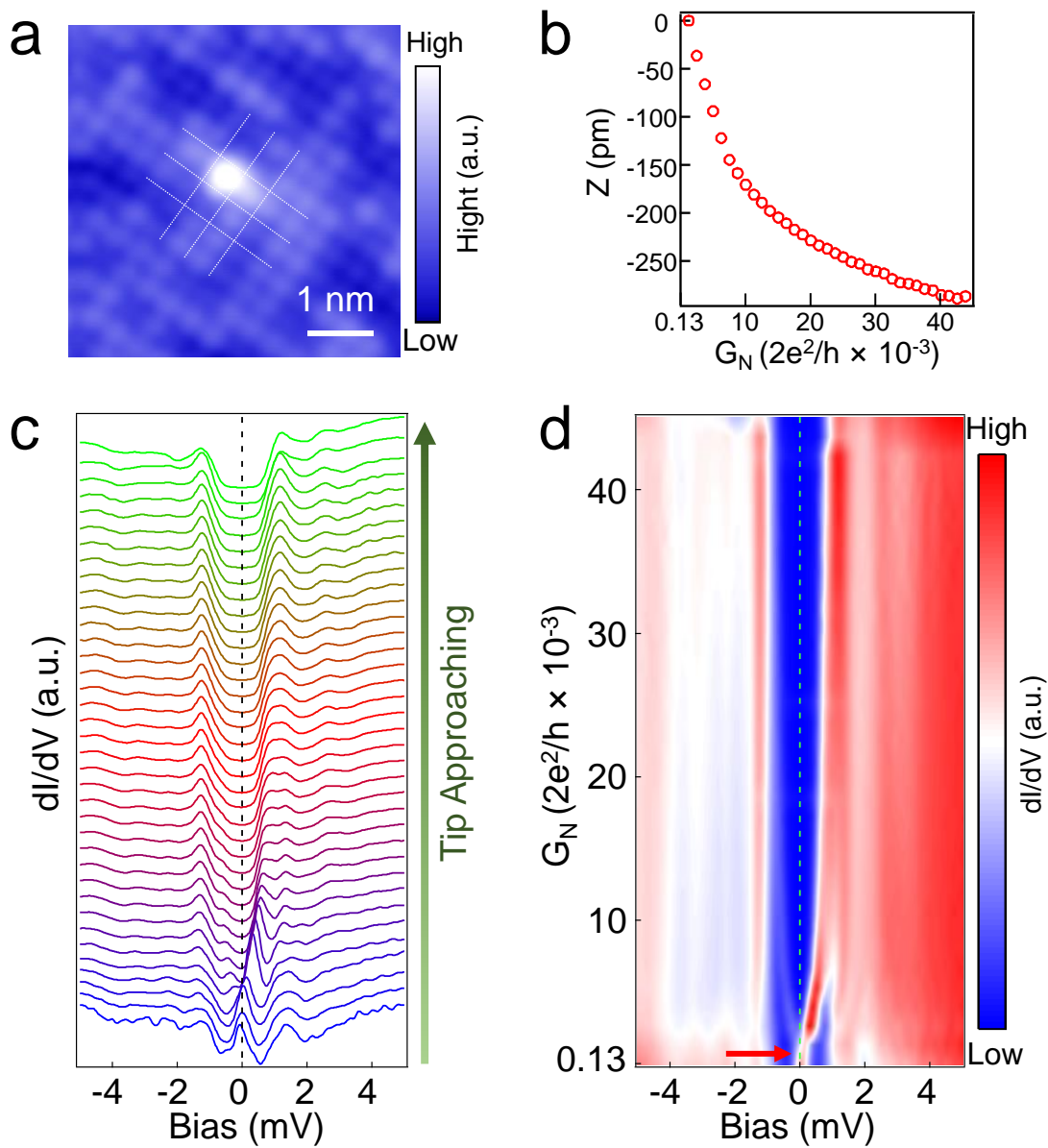

**Supplementary Figure 7. Modulating YSR states with approaching STM tip at a type-II Fe adatom.**
**a**, A high-resolution STM image ($V_s = -10$ mV, $I_t = 100$ pA) showing a single type II Fe adatom located away from the $C_4$ symmetric site (on the vertices of the white dashed grids). **b**, The tip-sample distance offset $z$ versus tunnel-barrier conductance $G_N$ as the tip approaches the adatom. The $z$-offset decreases smoothly with increasing barrier conductance $G_N$. **c** and **d**, The normalized dI/dV spectra ($V_s = -10$ mV, $I_t = 200$ pA) and the corresponding intensity plot under different tunnel-barrier conductance in zero applied magnetic field, showing the shift of YSR states. The red arrow indicates the quantum transition point of the YSR states.

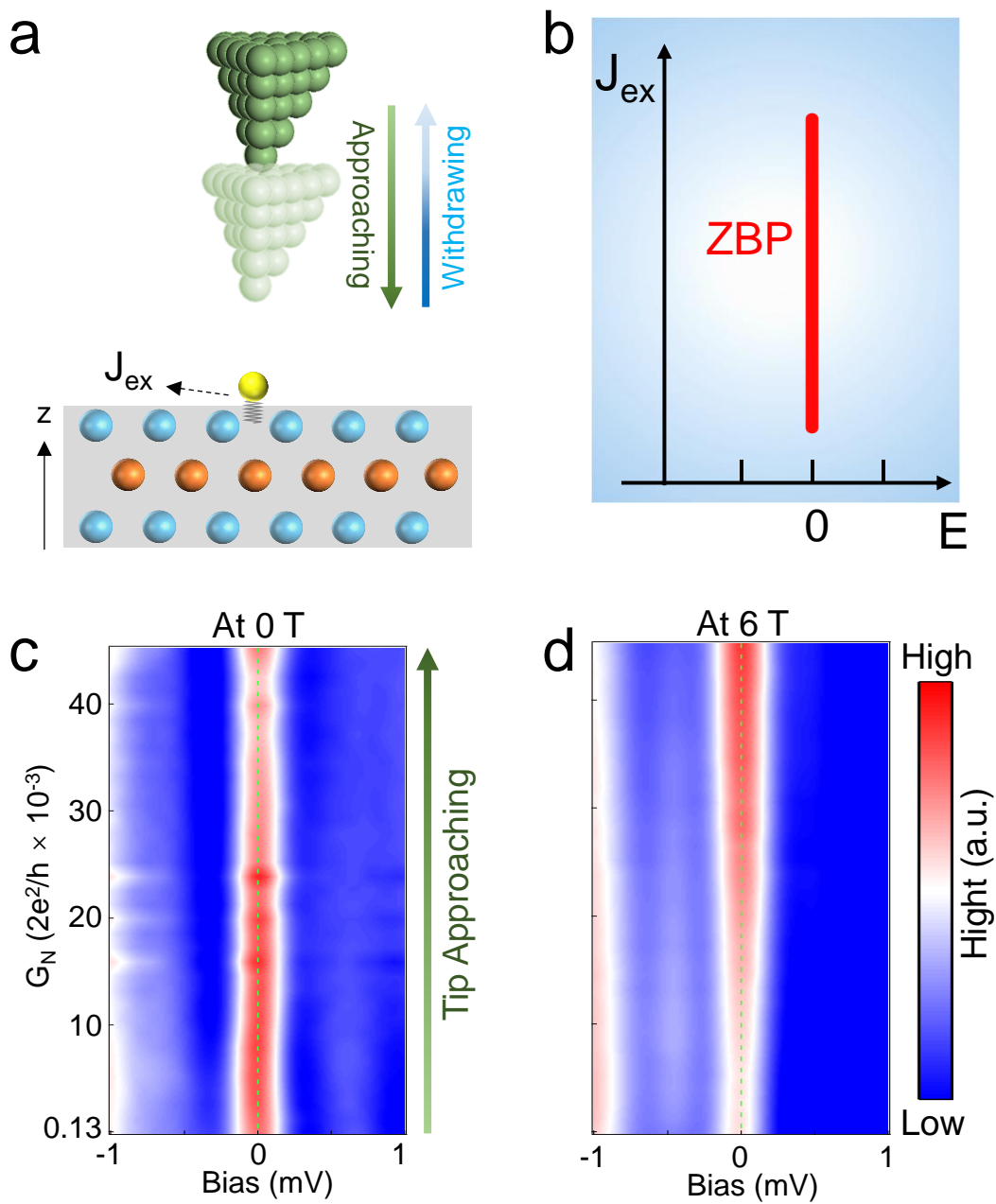

**Supplementary Figure 8. Robustness of ZBP under approaching STM tip at a type-I Fe adatom. a**, Schematics illustrating an approaching STM tip on top of a type-I Fe adatom. The Fe adatom is located at the high-symmetry site and is close enough to the surface. The exchange coupling is strong enough to produce a ZBP without approaching tip. **b**, Schematic diagram showing the robustness of the ZBP at type-I Fe adatoms against increasing exchange interaction ($J_{ex}$). **c** and **d**, Intensity plots of the dI/dV spectra, showing the evolution of the robust ZBP with increasing tunnel barrier conductance at 0 T (**c**) and 6 T (**d**).

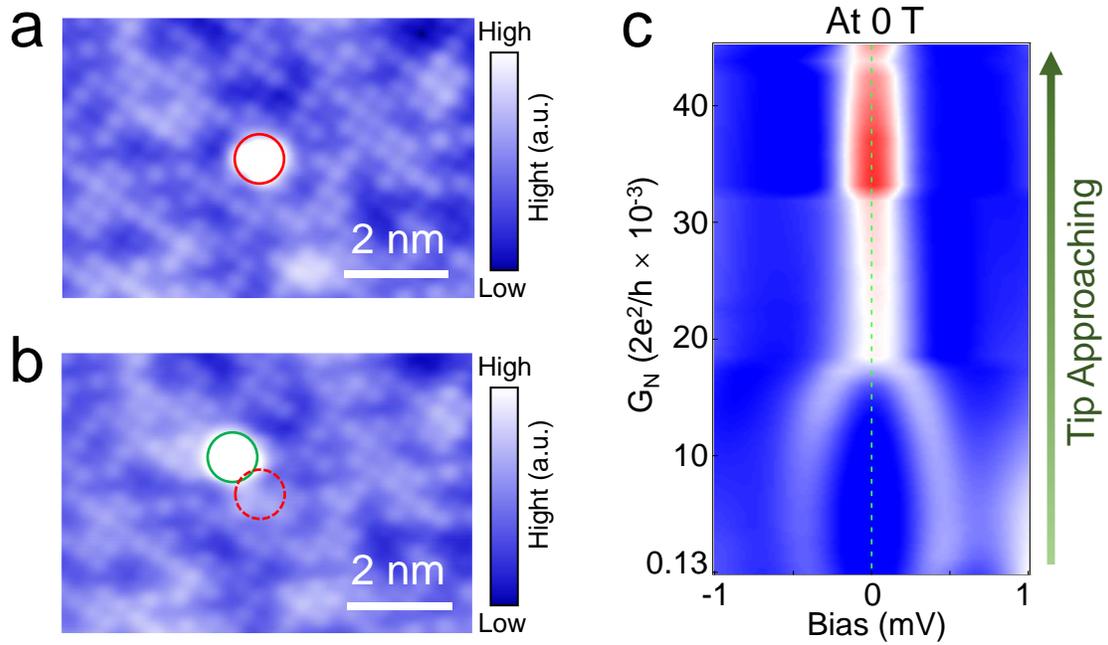

**Supplementary Figure 9. Replicable transition between YSR states and a vortex MZM. a** and **b**, STM images ($V_s$ = -10 mV, $I_t$ = 100 pA) showing the position of the Fe adatom before (red circle) and after (green circle) the STM manipulation. The dI/dV spectra in Figs. 3c-f of the main text correspond to the adatom located in the red circle. **c**, Intensity plot of dI/dV spectral evolution with tunnel barrier conductance at the Fe adatom located in the green dotted circle in **b**. The transition between the YSR states and the vortex MZM represented by the ZBP observed in Fig. 3 of the main text is reproduced at the new location.

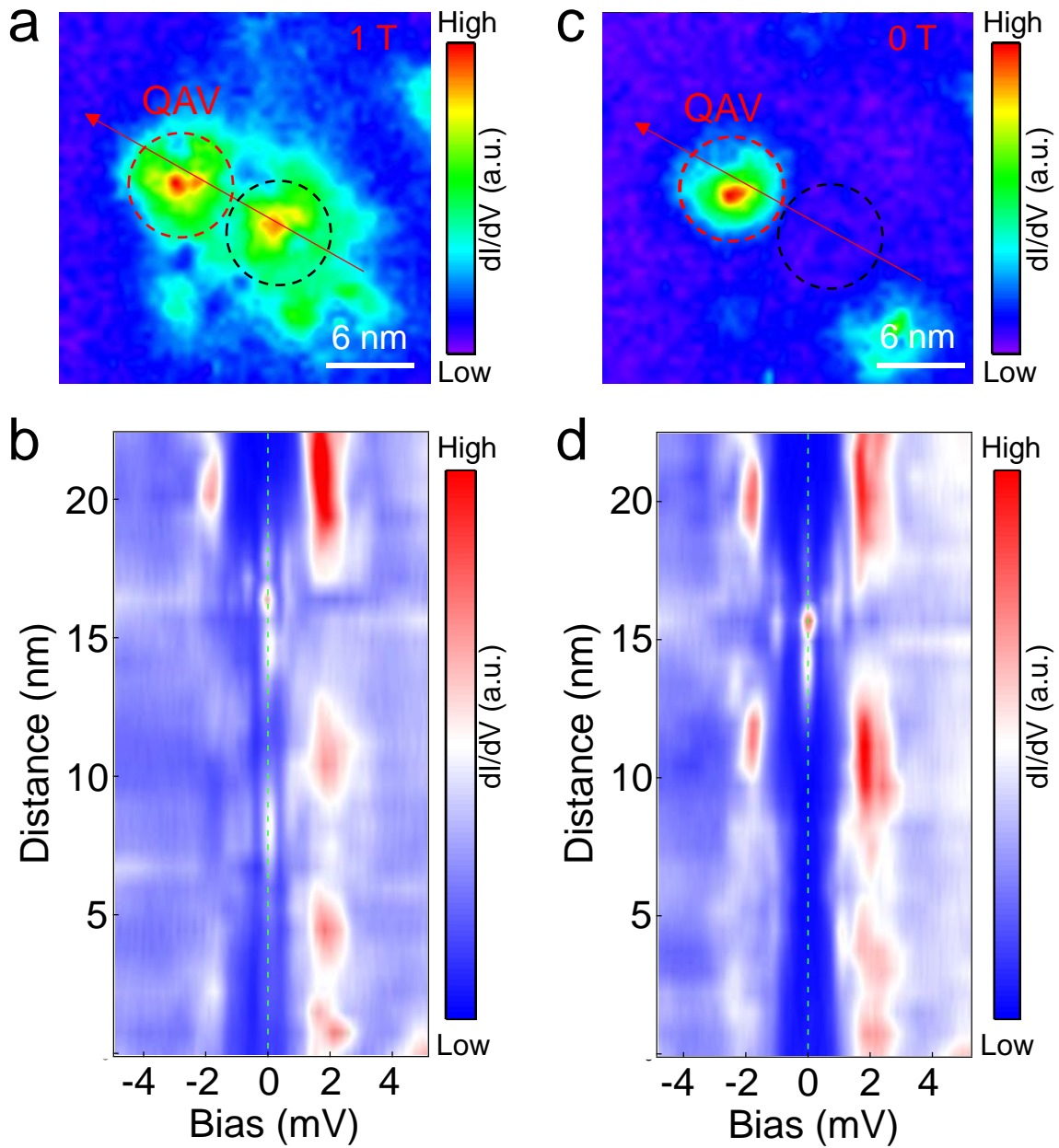

**Supplementary Figure 10. Coexistence of MZMs in a QAV and a field-induced vortex. a**, The zero-energy dI/dV map ($V_s = -10$ mV, $I_t = 100$ pA) in an applied field of 1 T around a type-I Fe atom. The QAV nucleated at the adatom (red dashed circle) is the same one shown in Fig. 4 of the main text. The center of the field-induced topological vortex (black dashed circle) is about 7 nm away from the center of the QAV. **b**, Intensity plot of dI/dV spectra ($V_s = -10$ mV, $I_t = 200$ pA) along the red arrow in **a**, showing two ZBPs localized separately in the cores of both vortices. **c**, The zero-energy dI/dV map ($V_s = -10$ mV, $I_t = 100$ pA) before applying the magnetic field in the same area. **d**, Intensity plot of dI/dV spectra ($V_s = -10$

mV, $I_t$ = 200 pA) along the red arrow in **c**, showing the hard superconducting gap in the region occupied by the field-induced vortex (black dashed circle).